\documentclass[twocolumn,prb,aps,showpacs]{revtex4}
\usepackage{graphicx}
\usepackage{dcolumn}
\usepackage{bm}

\begin{document}

\title{Tomographic-like reconstruction of the percolation cluster as a phase transition}

\author{Nira Shimoni}
\author{Doron Azulai}
\author{Isaac Balberg}
\author{Oded Millo}
\email{milode@vms.huji.ac.il} \affiliation{Racah Institute of
Physics, The Hebrew University, Jerusalem 91904, Israel}

\begin{abstract}
  We observed a phase transition-like behavior that is marked by the onset of the realization
  of the connectivity between two sites on a two-dimensional cross-section of a three-dimensional
  percolation cluster. This was found using contact-resistance atomic force microscopy on carbon
  black/polymer composites. The features in the current images, when presented as a function
  of the cut-off current, or as a function of the total area covered by the electrically
  connected objects, appear to obey a cluster statistics that is similar to the one predicted
  and observed in continuum percolation systems.
\end{abstract}
 \maketitle

While the basic theory of percolation seems to be well
understood,\cite{1} there are many aspects of classical
percolation theory that have not been investigated intensively in
spite of their fundamental importance and practical applications.
In particular, there is a group of problems associated with the
fact that usually one observes three-dimensional (3D) structures
with methods that yield two-dimensional images. Such are the
projection,\cite{2} the illumination \cite{3} and the
cross-section \cite{4} of 3D percolation systems. Conspicuous
applications are astrophysical maps \cite{5}, biological \cite{6}
and medical scans,\cite{7} meteorological \cite{3,8} and
hydrological \cite{9,10} information, and the determination of the
structure of the percolating phase in binary inhomogeneous
media.\cite{4,10,11,12,13}  In particular, it was only very
recently that the dynamic properties associated with these
problems have been considered.\cite{10} This was the phenomenon of
flow between two surface sites, that are connected by the
underlying 3D percolation cluster. In this paper we report a
manifestation of the above-mentioned problems in binary composites
using an experimental method that depends on the dynamical aspects
of flow within a percolation cluster. As we show below, the
connection between the geometrical and dynamical aspects can be
described by a percolation-type phase-transition-like behavior.

While manifestations of 3D properties in 2D data was found a long
time ago\cite{13} (on composites in which the electrical
properties have indicated a percolation behavior\cite{14,15}) the
more recent development of atomic force microscopy (AFM) and its
extensions, to stiffness microscopy,\cite{4} electrical force
microscopy (EFM) \cite{11} and contact-resistance microscopy
\cite{12} (C-AFM), have enabled more direct cross-section images
of the percolation cluster.  In particular, EFM images have been
used for the experimental determination of the fractal
dimension,\cite{16} $D_{f}$, of the 3D "percolating" conducting
phase in composites made of carbon black (CB) particles embedded
in an insulating polymer.\cite{11}

The initial aim of our study was to reexamine the above EFM
results using the superior \cite{12} C-AFM method, since the
results obtained in Ref. 11 were subjected to a significant
uncertainty yielding \cite{17} a $D_{f}$ value in the range 2.6
$\pm$ 0.6.  This permits a $D_{f}=3$ value, which is in the
homogeneous (3D) regime, and thus may shed doubt as to the
validity of the underlying scaling assumption claimed in Ref. 11
(see below).  The latter should yield the value borne out from
percolation theory,\cite{1,11} $D_{f}=2.53$.

Our samples, with CB fillings between 10 and 23 vol\%, were
prepared from a commercial grade polyethylene and Vulcan XC-72
carbon black \cite{15}. This CB has 100-1000 nm-long aggregates
that consist of fused spherical graphite particles that are 30 nm
in diameter \cite{14,18}. The composites were prepared \cite{11}
in a common compounding and molding procedure, \cite{19,20}
resulting in uniform plaques 0.25 mm thick. The plaques were cut
to long strips for the macroscopic measurements, described
previously, \cite{15} and to squares of 12x12 mm$^{2}$ for the
present C-AFM study. A 5 mm wide silver-paint strip at the sample
side served as a counter electrode to the local contact provided
by the conducting tip. The C-AFM images were acquired more than 2
mm away from the strip, thus the current between the tip and the
electrode is possible only if there is an "infinite" conducting
cluster in the system, i.e., the current must flow between the tip
and a branch of this infinite cluster.  We call a cross section of
such a branch, as reflected in the surface image, a "conducting
island".
\begin{figure*}
    \includegraphics[width=12cm]{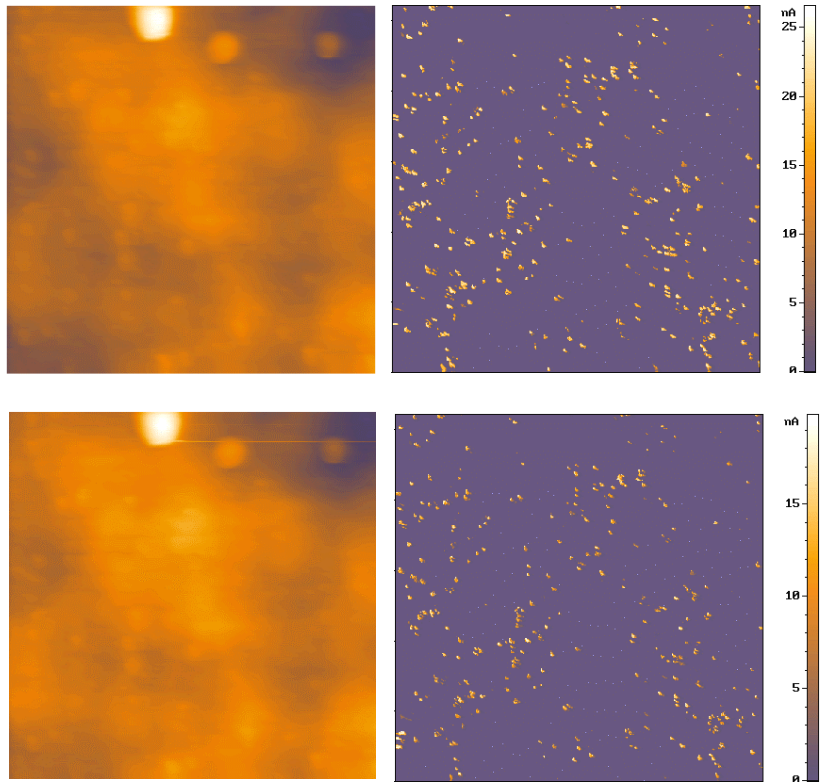}
    \caption{5x5 $\mu \textrm m^{2}$ topographic  (left) and current
(right) images for a sample of 15 vol\% CB filling, obtained for
two values of applied voltage, $V = 1.8$ V (lower scan) and $V =
2.8$ V (upper scan).  The color scale range in the topographic
images is 200 nm, while in the current images it is 20 nA and 27
nA for the 1.8 V and 2.8 V scans, respectively.} \label{fig1}
\end{figure*}

Our measurements were performed using a commercial (NT-MDT Solver)
Scanning Force Microscope in the constant-force mode, with 0.03
N/m cantilever stiffness. The current maps were measured along
with the topography for a given tip-electrode voltage, $V$.  The
silicon cantilever tips had curvatures of less than 35 nm, and
were coated by a 25 nm-thick titanium-nitride film.  We thus
estimate the resolution in the topographic images to be better
than 60 nm. The spatial resolution of the current maps is,
however, much better, and was limited by our pixel size, which is
of the order of 10x10 nm$^{2}$.  This is important considering the
size of the conducting particles (the CB aggregates, see above),
since this means that we can assume that a pixel in an image
belongs to a single particle. The noise level was around 0.02 nA
in most measurements and thus in our image analysis we did not
consider currents below 0.04 nA.  The images discriminate between
different current intensities (between and within islands) and
thus enable statistical analysis of the current images taken with
a constant voltage. In particular, we extract from each image the
individual and inclusive areas of the islands for currents larger
than a chosen cutoff current $I_{\textrm{co}}$.

Turning to the results, in Fig. 1 we show typical topographic and
current images obtained by the C-AFM for two values of $V$.  We
found no correlation between the surface morphology and the
distribution of the "conducting islands". This implies that the
current maps reflect mainly the bulk sample properties, namely,
the 3D percolation cluster. The increase of the number of islands
and their individual areas with $V$ underlines the 3D nature of
the system as more percolation cluster branches can be detected
with increasing $V$. The increase of the bias appears then to be
associated with "deeper" cross sectioning of the conducting
network of the bulk.

\begin{figure}
    \includegraphics[width=8cm]{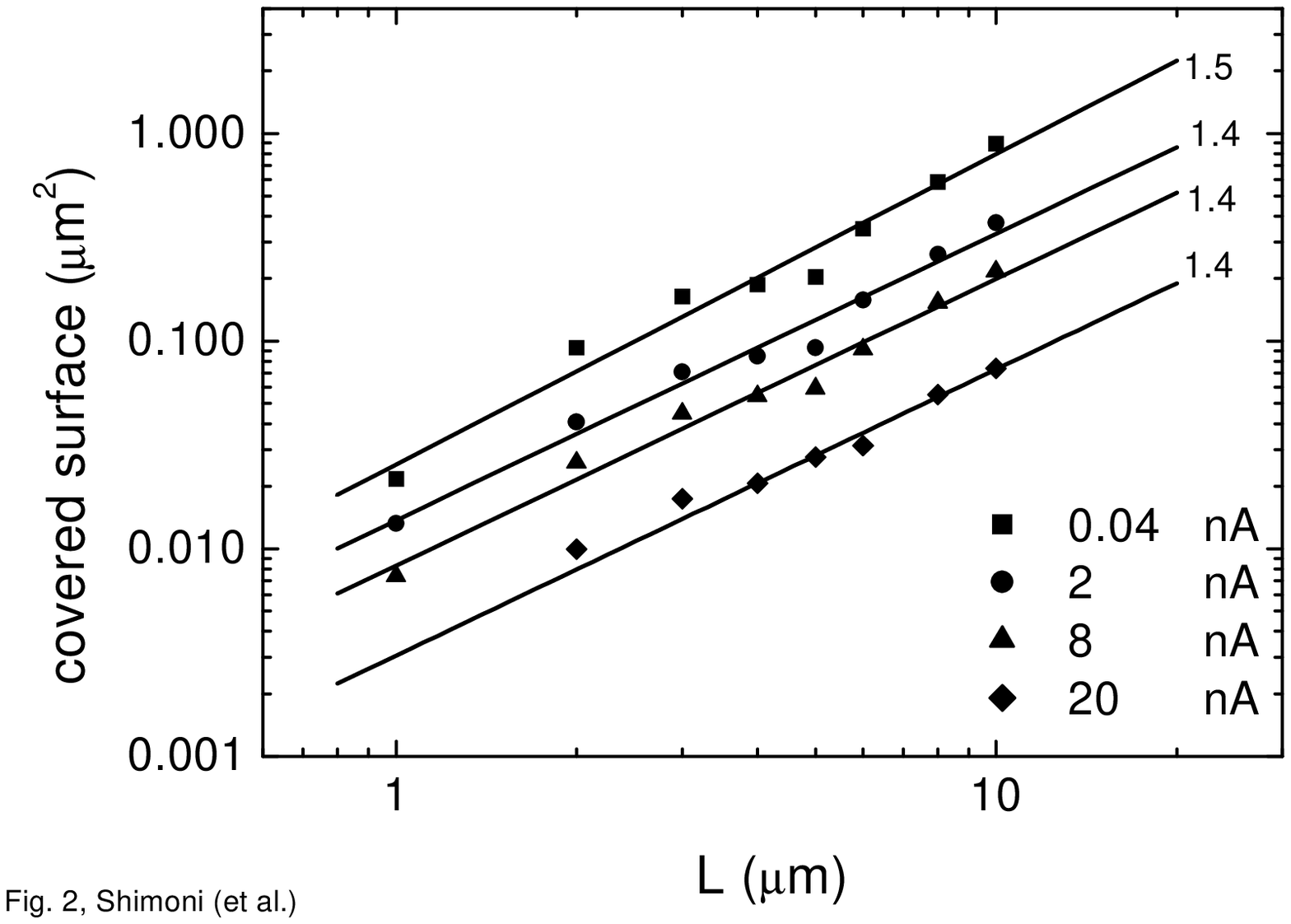}
    \caption{The measured total area covered by "conducting islands",
$A_{t}$, as a function of the square window size $L$ for a sample
with 10 vol\% CB filling, for four different cutoff currents. The
lines are fits to the scaling dependence, $A_{t} \sim L^{D_s}$,
with the indicated $D_{s}$ values. The voltage applied in this
case was 2V. The points in this and in other graphs represent mean
values of data obtained for various samples and scans.}
\label{fig2}
\end{figure}
To check the fractal dimension of the islands' area on the
surface, $D_{s}$, we first considered samples with different CB
contents. We measured the area covered by all the islands,
$A_{t}$, as a function of the scanned area ("the window"),
$L^{2}$. This was done first in the regime for which scaling
behavior is expected,\cite{1} i.e., $\nu$ = 10 vol\%, which is
close to the percolation threshold, $\nu_{c}$ (= 9.3 vol\%)
\cite{15}, and $L$ larger than the particle size ($\sim$ 0.5
$\mu\textrm m$ in our case) but smaller than the correlation
length (which is of macroscopic order near the percolation
threshold, \cite{12} say, larger than 10 $\mu \textrm{m}$). The
results of our measurements for various cutoff currents are shown
in Fig. 2.

For all nine cutoff currents considered (in the range $0.04\leq
I_{\textrm{co}} \leq 20 $ nA), the relation $A_{t}\propto
L^{D_{s}}$ (expected from percolation theory \cite{1}) was obeyed,
as seen in Fig. 2 for four cutoff currents. The extracted exponent
for each of the nine cutoff currents has a value within the
interval $1.35\leq D_{s} \leq 1.65$.  Our $D_{s} = 1.5 \pm 0.15$
exponent has the same value as that obtained \cite{11,17} from EFM
measurements, but it has, as explained above, a higher certainty
which is very significant. Utilizing the expectation \cite{11,16}
that the fractal dimension of the bulk cluster is given by $D_{f}
= D_{s}+1$, we find that $D_{f} = 2.50 \pm 0.15$, in agreement
with the value expected \cite{1} for a 3D percolation cluster,
2.53. We have further established the validity of the above
values, in addition to the smaller error bars and the fact that
they were found for nine different cutoff currents, by an even
more significant test. This is by performing the same measurements
on a sample with $\nu$ well removed from $\nu_{c}$, where the
"normal" 3D dimensionality should be recovered.  Three cutoff
currents (0.5, 4, and 20 nA) were used in a measurement of a $\nu
= 15$ vol\% sample, all yielding a value of $D_{s} = 2.00 \pm
0.05$. The clear distinction between the results for the 10 vol\%,
and the 15 vol\% samples is very convincing evidence that our
surface measurements reveal properties of the bulk percolation
cluster. This behavior appears to be general, as it was found not
only for the present "high-structure" \cite{14,15,18} CB/polymer
composite but also for another, "low-structure", CB/polymer
composite that we have studied \cite{15}, as well as for granular
Ni/SiO$_2$ films. The CB system, however, is of particular
interest, since it is not a "true" percolation system, in the
sense that conduction takes place via tunneling between the CB
particles even in the percolation cluster. Nevertheless, it does
conform with 3D continuum percolation behavior.\cite{21}

The results shown in Fig. 2 have significant implications beyond
the latter conclusion.  We see that, systematically, the lower the
value of $I_{\textrm {co}}$ the larger the value of $A_{t}$, for
the same $V$ and $L$. Let us consider this significance.  All the
conducting islands on the surface belong to the percolation
cluster, as otherwise, no current can flow between them and the
counter-electrode.  On the other hand, the value of $I_{\textrm
{co}}$ determines the corresponding longest possible current paths
that can participate in the conduction (see below). We can then
interpret the values of $I_{\textrm {co}}$ in terms of
connectivity. The higher the $I_{\textrm {co}}$, the smaller the
ensemble of conductance paths between the observed islands and the
electrode via the percolation cluster. But then, two islands that
are seen at a given $I_{\textrm {co}}$ are also \emph{connected
among themselves} via the 3D percolation cluster. The lower the
value of $I_{\textrm {co}}$, the longer the conductance paths
included in the ensemble of possible paths and the larger the
number of detected "conducting islands" that are interconnected
via the percolation cluster.  Thus the variation of $I_{\textrm
co}$ acts as an electrical cross-sectioning parameter of the
\emph{3D} percolation cluster. In other words, the lower the value
of $I_{\textrm {co}}$, the "deeper" the (electrically) added
"tomographic-like" section of this cluster. So, while it is
obvious that all the conducting islands are connected, the
electrical \emph{realization} of this connectedness is limited in
our measurements to part of the conductance paths in the system.
This interpretation gives a meaning to the $I_{\textrm {co}}
\rightarrow 0$ limit as the $I_{\textrm {co}}$ value for which all
available connecting paths between two islands are realized.  We
have then a \emph{transition} from a \emph{partially realized}
(via the 3D percolation cluster) connectivity of the islands in
the system to a \emph{fully realized} connectivity of this system.
Our experimental approach here is closely related to the
well-known concepts of optimal and shortest paths.\cite{22} In
fact, the higher the chosen current cutoff, the closer the
collection of current paths to the shortest electrical (and
statistically, probably also the geometrical "chemical distance"
\cite{22}) path. Our $I_\textrm{co}$ values then represent a
hierarchy of collection of paths that deviate from the shortest
electrical path. The $I_{\textrm {co}} \rightarrow 0$ limit is
associated with the collection of all possible paths.

\begin{figure}
    \includegraphics[width=8cm]{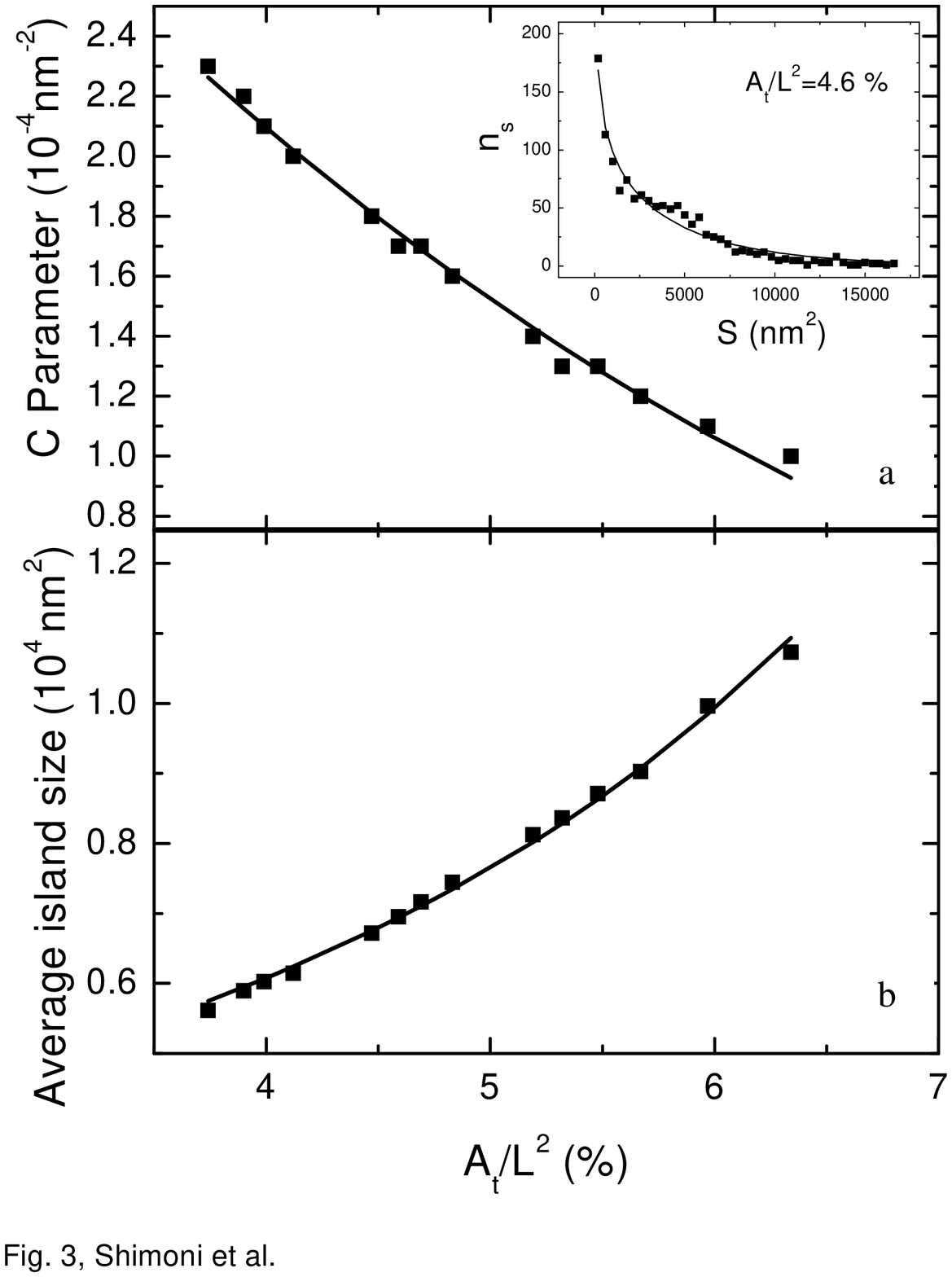}
    \caption{ (a) The dependence of the parameter $C$ on the
fractional total covered area, $A_{t}/L^{2}$. Inset: the number of
observed islands as a function of their size for $A_{t}/L^{2} =
4.6 \%$. (b) The dependence of the average size of the observed
"conducting island" on the fractional total covered area,
$A_{t}/L^{2}$. The measurements were carried out for $\nu = 15$
vol\%, $L = 10 \mu \textrm m$ and $V = 2$V. The lines present fits
to percolation theory-like expressions, see text. }\label{fig3}
\end{figure}

Can the above onset of realized connectivity at $I_{\textrm {co}}
\rightarrow 0$ be described as a percolation-type phase
transition? A priori there is a difficulty in trying to examine
this question, since we do not know the distribution of
conductance paths in order to apply $I_{\textrm {co}}$ as a
"thermodynamic parameter" \cite{23} of a phase transition.
However, there is an obvious parameter that $I_{\textrm {co}}$ can
be mapped to (see Fig. 2) and with which we are familiar from the
theory of continuum percolation \cite{21,22,24}. This parameter is
the fractional "covered" or "occupied" area $A_{t}/L^{2}$ that can
be simply mapped onto the site occupation probability $p$ in
lattice percolation \cite{21}. The choice of $A_{t}/L^{2}$ as a
"thermodynamic parameter" that describes the cross sectioning
mentioned above also seems to suggest the value of its critical
point, $A_{tc}/L^{2}$, as the value of $A_{t}/L^{2}$ when
$I_{\textrm {co}} \rightarrow 0$. We point out again that the
value of $A_{tc}/L^{2}$ does not yield any particular feature in
the observed image, and that the transition is "hidden" here by
the underlying 3D percolation cluster.  The parameter
$(A_{t}-A_{tc})/L^{2}$ may then be, as in lattice or continuum
percolation, the parameter that characterizes the proximity to the
critical point \cite{21,24}. Having this parameter, it was natural
to determine if cluster statistics such as in phase transitions in
general,\cite{23} and in continuum percolation in
particular,\cite{1,22,24} will be observed in the present case.
For that purpose we examined the behavior of two quantities, that
can be readily monitored and analyzed. To find whether they show
the behavior of a "phase transition" we have to check if they obey
the corresponding power-law behaviors\cite{1,22,23} for the
\emph{same} $A_{tc}$.

Following the above considerations we have plotted the number of
observed islands, $n_{s}$, as a function of the island area, $S$,
and we found that, as shown in the inset of Fig. 3(a) for
$A_{t}/L^{2}=4.6$ \%, the results can be well fitted by a $n_{s}
\propto S^{-\tau}\exp(-CS)$ behavior (as in percolation theory
\cite{1,22}). The specific prediction of percolation theory is
that $C \propto (A_{t}-A_{tc})^ {1/\sigma}$ where $\tau$ and
$1/\sigma$ are dimensional-dependent exponents. As shown in Fig.
3(a), very good fits to the data with that dependence are achieved
for a sample of $\nu = 15$ vol\%. The best fit yielded
$A_{tc}/L^{2} = 13$\%, $\tau = 0.25$ and $1/\sigma = 2.65$. An
analysis of different areas of this same sample yielded
$A_{tc}/L^{2}$ values in the range of $13 \pm 1$ \ 
results were obtained for other values of $\nu$, and the
corresponding values of $A_{tc}$ were found to increase with
$\nu$.

The other dependence that we evaluated was that of the average
size of the conducting islands (the cluster size in percolation
theory \cite{1,22}) $<S>$ on the above proximity parameter. $<S>$
is defined in percolation theory as $\Sigma n_{s}S^{2}/\Sigma
n_{s}S$, and is predicted to behave as $(A_{t}-A_{tc})^{-\gamma}$.
We checked then whether the data obtained for $<S>$ can be fitted
by that power-law. The best fit for the 15 vol\% sample yielded
$\gamma = 1.9$ and, \emph{again}, $A_{tc}/L^{2} = 13$\%. This is
shown in Fig. 3(b).

The important conclusion that can be derived from Fig. 3 is that
the dependencies of $C$ and $<S>$ on $(A_{t}-A_{tc})$ can be
fitted very well to a power law, both with the same $A_{tc}$ and
with percolation theory-like exponents. The quality of the data is
not sufficient for obtaining exact values of the exponents, but
the fact that the \emph{same} $A_{tc}$ value is found with
\emph{reasonable exponents} does establish that we have here a
behavior that can be well described as a phase transition. The
interesting finding here is that while we do not have any
"visible" phase transition in the image (such as, say, 2D
connectivity), the quantity $A_{t}$ (or the hidden variable
$I_{\textrm {co}}$) serves as the "thermodynamic parameter" of the
phase transition in the \emph{realization of the connectivity} of
the underlying 3D system. The dynamic aspects, emphasized here by
the dependence of the connectivity on $V$, are similar to the
conduction under constant pressure difference in the case of flow
between two sites in porous media.\cite{10} In fact, as will be
described elsewhere, the values of $A_{t}$ that are determined for
different values of $V$ exhibit a similar behavior to the one
described here for different values of $I_{\textrm {co}}$.

\acknowledgments { The authors would like to thank M. B. Heaney
and M. Wartenberg for the samples used in this study and A.
Aharony, K. Schwartz and A. Drory for helpful discussions. This
work was supported by the Israel Science Foundation and the
Niedersachsen Foundation.}

\end{document}